\documentclass[prd,preprintnumbers,floatfix,
nofootinbib,superscriptaddress]{revtex4}

\usepackage{float}
\usepackage{nicefrac}
\usepackage{mathtools}
\usepackage{amsfonts} % AMS
\usepackage{amssymb} % AMS
\usepackage{amsmath} % AMS
\usepackage{bbold}
\usepackage{bm} % bold math
\usepackage{graphicx} % Include figure files
\usepackage{subfigure} % Include figure files
\usepackage{array} % array
\usepackage{dcolumn} % Align table columns on decimal point
\usepackage{bm} % bold math
\usepackage{latexsym} % latex symbols
\usepackage{longtable} % long tables
\usepackage{hyperref} % hypertext links 
\usepackage{verbatim}
\usepackage{epsfig}
\usepackage{slashed}
\usepackage{color}

\usepackage{glossaries-extra}
\setabbreviationstyle[acronym]{long-short}
\glssetcategoryattribute{acronym}{nohyperfirst}{true}

\newcommand{\cA}[0]{\mathcal A}

\newcommand{\cL}[0]{\mathcal L}

\newcommand{\cO}[0]{\mathcal O}

\newcommand{\cV}[0]{\mathcal V}

\newcommand{\wt}[0]{\widetilde}

\newcommand{\chpt}{$\chi$PT}
\newcommand{\SU}{{\rm SU}}
\newcommand{\str}{{\rm str}\!}

\newcommand{\DSPQLuscher}[0]{Draper:2021clv}
\newcommand{\HSconstraint}[0]{Hansen:2011kk}
\newcommand{\HSWChPT}[0]{Hansen:2011mc}

\newcommand{\Luscher}[0]{Luscher:1986pf,Luscher:1990ux}

\newacronym{CMF}{CMF}{center-of-momentum frame}

\definecolor{darkgreen}{rgb}{0,0.8,0}

\begin{document}

%\preprint{\vbox{\hbox{JLAB-THY-19-2945} }}

\title{$\pi\pi$ scattering in partially-quenched twisted-mass chiral perturbation theory}

%%%%%%%%%%
\author{Zachary T. Draper}
\email[e-mail: ]{ztd@uw.edu}
\affiliation{Physics Department, University of Washington, Seattle, WA 98195-1560, USA}

%%%%%%%%%%
\author{Stephen R. Sharpe}
\email[e-mail: ]{srsharpe@uw.edu}
\affiliation{Physics Department, University of Washington, Seattle, WA 98195-1560, USA}
%%%%%%%%%%

%%%%%%%%%%

%%%%%%%%%%
\date{\today}
%%%%%%%%%%

%%%%%%%%%%
\begin{abstract}
We study pion-pion scattering in partially-quenched twisted-mass lattice QCD
using chiral perturbation theory. The specific partially-quenched setup corresponds to
that used in numerical lattice QCD calculations of the $I=0$ scattering length.
We study the discretization errors proportional to $a^2$, with $a$ the lattice spacing,
and the errors that arise due to the use of L\"uscher’s two-particle quantization condition in a theory
that is not unitary. We argue that the former can be as large as $\sim 100\%$, but
explain how they can be systematically subtracted using a calculation of the
$I=2$ scattering amplitude in the same partially-quenched framework.
We estimate the error from the violation of unitarity to be $\sim 25\%$, and argue that
this error will be difficult to reduce in practice.
 \end{abstract}
%%%%%%%%%%

%\keywords{weak decays, lattice QCD}

\nopagebreak

\maketitle

%%%%%%%%%%%%%%%%%%%%%%%%%%%%%%%%%%%%%%%
%%%%%%%%%                          PAPER                             %%%%%%%%%
%%%%%%%%%%%%%%%%%%%%%%%%%%%%%%%%%%%%%%%

\section{Introduction}
\label{sec:intro}

%\input{intro}
% !TEX root = twoplusone.tex

% This is intro.tex
% \label{sec:intro}
%

Considerable progress has been made over the last decade in studying two-particle
scattering at physical, or near-physical, quark masses~\cite{Fu:2013ffa,Sasaki:2013vxa,ETM:2015bzg,Liu:2016cba,Fu:2017apw,Culver:2019qtx,Mai:2019pqr,Fischer:2020jzp,Fischer:2020yvw}.
Our focus here will be on calculations 
using the twisted-mass (TM) formulation of
lattice QCD (LQCD)~\cite{Frezzotti:2000nk}.
In particular, the ETM collaboration has used TM fermions 
to determine scattering lengths in the following channels: 
$K^+K^+$~\cite{Helmes:2017smr},
$K^+\pi^+$~\cite{Helmes:2018nug},
and $\pi\pi$ scattering for $I=2$~\cite{Feng:2009ij,ETM:2015bzg}
and $I=0$~\cite{Liu:2016cba}.
All these works are done at maximal twist,
and make use of the two-particle quantization condition derived by L\"uscher~\cite{\Luscher}.

The work we consider in most detail here is the calculation of the 
$I=0$ $\pi\pi$ scattering length in Ref.~\cite{Liu:2016cba}.
The $I=0$ channel is the most
challenging from a numerical point of view, due to the presence
of fully disconnected quark diagrams, 
and is likely to have the largest systematic errors.
For technical reasons described below, Ref.~\cite{Liu:2016cba}
used a partially quenched (PQ) setup,
in which some of the valence quarks have differing twist angles from the corresponding sea quarks.
These valence quarks are referred to as Osterwalder-Seiler (OS) quarks~\cite{Frezzotti:2004wz}.
Using a PQ theory implies that unitarity is violated~\cite{Bernard:1993sv,Bernard:2013kwa}, 
which introduces an additional source of systematic error.
Our aim here is to use chiral perturbation theory (\chpt) to shed new light on the errors due to
discretization and the violation of unitarity.
%so as to help control the extrapolation to the continuum limit in future work.
 In this regard we note that, so far, results are only available at a single lattice spacing.
The specific version of \chpt\ that we use incorporates the effects
of partial quenching and discretization errors due to the twisted-mass,
and is denoted PQTM\chpt.

A key feature of TMLQCD at maximal twist is that discretization errors 
for physical quantities are automatically $\cO(a)$ improved~\cite{Frezzotti:2003ni}.
Even so, when working at the physical point,
generic discretization effects are comparable to those due to the nonzero values of the physical 
light quark masses. To see this, recall that in \chpt,
a measure of the former effects is $a^2 \Lambda_{\rm QCD}^2$, while that of the latter
is $M_\pi^2/\Lambda_{\rm QCD}^2 \approx m_q/\Lambda_{\rm QCD}$.
Now, if $m_q\approx 5\;$MeV, $1/a\approx 2\;$GeV, and $\Lambda_{\rm QCD}\approx 300\;$MeV,
 then $m_q/\Lambda_{\rm QCD} \approx 0.016$, 
while $a^2 \Lambda_{\rm QCD}^2 \approx 0.022$, which is indeed of the same size.
A concrete example of this result (to be discussed further below) is that a pion composed of OS quarks
has, at present lattice spacings,
a mass of about 250 MeV when the corresponding unquenched pions have masses close
to 135 MeV. 
This difference is an $\cO(a^2)$ effect, which is seen to be as large as the pion mass-squared
itself.
This implies that there can be large discretization errors in quantities
that are proportional to $M_\pi^2$, 
such as the pion scattering amplitudes near threshold.

One goal of this work is to determine when such $\cO(a^2)$ errors are present.
In this regard, we note that Ref.~\cite{Buchoff:2008hh}  found, using TM\chpt, 
that there are no such errors for $I=2$ scattering with {\em unquenched} TM fermions.
Instead, the leading discretization errors are proportional to $a^2 m_q$, and thus suppressed by $a^2\Lambda_{\rm QCD}^2$ compared to the physical result.
This work also showed, however, that studying $I=0$ scattering with unquenched TM fermions
was challenging from a theoretical point of view because of mixing with $I=2$, $I_z=0$ states,
due to the breaking of isospin. 
Specifically, they found that this mixing occurred at $\cO(a^2)$, and thus was
comparable in magnitude to the scattering amplitudes themselves at the physical point.
Because of this problem, 
Ref.~\cite{Liu:2016cba} used pions composed of OS quarks to study $I=0$ $\pi\pi$
scattering, since there is then an exact valence isospin symmetry, 
and the above-described mixing with the $I=2$ channel does not occur.
This setup has not been studied previously using \chpt, 
and we carry out the calculation here in order to determine the form of the leading
discretization errors in this approach.

The second issue we address is the use of the two-particle quantization condition
in Ref.~\cite{Liu:2016cba}.
As we have stressed recently in Ref.~\cite{\DSPQLuscher},
and as is discussed in Ref.~\cite{Chen:2005ab},
unitarity is needed in order to use this quantization condition.
Since PQ theories are not unitary, in general one cannot use the quantization condition
to determine PQ amplitudes.
However, we argued in Ref.~\cite{\DSPQLuscher} 
that the quantization condition was likely valid if the lack of unitarity
was restricted to the $t$ and $u$ channels, i.e. if unitarity in the $s$ channel still held.
We also proposed alternative versions of the criteria that must be satisfied for a two-particle
channel in a PQ theory to be ``physical enough'' to allow use of the quantization condition.
Here we apply these criteria to the case of OS pion scattering.

\begin{comment}
The diagrammatic version of the criterion required studying the quark-line 
diagrams~\cite{Sharpe:1992ft} and seeing if $s$-channel loops of sea quarks contributed---if so,
 then the quantization condition would be invalid. As we explain,
the arguments of Ref.~\cite{\DSPQLuscher} show that the $I=2$ amplitude 
calculated with OS quarks does satisfy this criterion,
while the corresponding with $I=0$ and $I=1$ amplitudes do not. 
This version of the criterion is not, however, conclusive, as there could be cancellations between
unphysical intermediate states. 
The issue can be settled by an explicit calculation of the one-loop $s$-channel diagrams in
PQTM\chpt, and this is the second main result we provide in this paper.
\end{comment}

The remainder of this paper is organized as follows.
Section~\ref{sec:setup} describes the set-up within TMPQ\chpt.
Section~\ref{sec:LO} presents the results of the leading order calculation of
OS pion masses and scattering amplitudes. 
Section~\ref{sec:loop} studies the unitarity of the amplitude at one-loop order
[which is of next-to-next-to-leading order in our power counting].
Section~\ref{sec:imp} discusses the implications for the
results presented in Ref.~\cite{Liu:2016cba}.
We conclude in Sec.~\ref{sec:conc}.
Some technical results concerning the NLO contributions,
which we find do not give corrections to the scattering amplitudes or masses,
are relegated to Appendix~\ref{sec:NLO}.

\section{Theoretical setup}
\label{sec:setup}

In this section we construct the low-energy effective chiral theory that corresponds
to the calculations carried out in Ref.~\cite{Liu:2016cba}.
We note that this work used $N_f=2$ sea quarks.\footnote{%
Much recent work using TM fermions uses
$N_f=2+1+1$ sea quarks, with nondegenerate strange and charm sea quarks
(see, e.g. Ref.~\cite{ETM:2015bzg}).
Although a calculation of $I=0$ $\pi\pi$ scattering has not yet been attempted with
$N_f=2+1+1$ TM fermions, any future such calculation would still, at long distances, be described
by the effective chiral theory that we construct.
This is because, if we work at energies such that kaon-antikaon pairs cannot be created,
we can integrate out the strange (and charm) quarks leaving an effective lattice theory
with only up and down quarks, and having renormalized couplings.}

TMLQCD involves isodoublets of quarks, which have dimensionless mass terms of the form
\begin{equation}
m_0 + i  \mu_0 \gamma_5 \tau_3\,,
\label{eq:TM}
\end{equation}
as well as a standard Wilson derivative term and possibly an improvement (``clover'') 
term~\cite{Frezzotti:2000nk}.
We denote the doublet of unquenched (sea) quarks by $(u_S,d_S)$.
We say that $u_S$ has a positive twist and $d_S$ a negative twist,
because of the signs in the diagonal components of $\tau_3$.
To obtain maximal twist,
the ``normal'' bare lattice mass $m_0$ must be tuned to its critical value, 
such that the renormalized normal mass vanishes.
The entire mass of the quarks is then due to the bare lattice twisted mass, $\mu_0$.
The details of this procedure and its advantages are well 
known~\cite{Frezzotti:2003ni,Frezzotti:2004wz}
and we do not repeat them here.
We only note that the presence of the $\gamma_5\tau_3$ in the mass term leads to
explicit breaking of isospin and parity, although this vanishes in the continuum limit. 

As noted in the introduction, the isospin breaking introduced by the twist in the mass
creates mixing between the $I=0$ and $I=2, I_{z}=0$ states,
and Ref.~\cite{Liu:2016cba} avoids this using OS valence fermions.
To describe this setup in \chpt\, 
we introduce a doublet of OS valence fermions $(d_V,u_V)$,
along with a corresponding ghost-quark doublet $(\wt d_V,\wt u_V)$,
with both new doublets having the same mass term as in Eq.~(\ref{eq:TM}).
The ghost quarks (``ghosts'' for short)  cancel the determinant produced by the
valence quarks~\cite{Bernard:1993sv}.
We have interchanged the order of the flavors relative to the sea-quark doublet $(u_S,d_S)$,
so that the signs of the twists for the two valence flavors are opposite to 
those of the corresponding sea quarks.
(Alternatively, this sign flip could be achieved by replacing $\tau_3$ with $-\tau_3$ in the mass term,
while leaving the order of the flavors in the doublet unchanged.)
Choosing the OS quarks to have opposite twist from the sea quarks of the same flavor
allows us to pick out doublets containing one up and one down quark with the same sign 
of the twist. In particular, the doublet
$(u_S,d_V)$ has positive twist, while $(u_V,d_S)$ has negative twist.
There is an exact $\SU(2)$ isospin symmetry acting on each doublet with definite twist,
so by constructing pions composed of quark-antiquark pairs from one of these doublets,
we avoid mixing between $I=0$ and $I=2, I_{z}=0$ states. 
In particular, we focus on $(u_S,d_V)$ and refer to the isospin symmetry of this doublet
 as $\SU(2)_+$ symmetry, with the subscript denoting positive twist.
In the following, we will sometimes refer to this pair as the ``OS quarks'',
and the $SU(2)_+$ symmetry as ``OS isospin.''

All six fields are collected into a single spin-$1/2$ field,
\begin{equation}
Q^{\rm Tr} = (u_S, d_S, d_V, u_V, \wt d_V, \wt u_V)\,.
\label{eq:Q}
\end{equation}
If the twisted part of the mass terms were absent, then the theory
would have an $\SU(4|2)$ graded flavor symmetry, which includes
a flavor $\SU(4)$ in the quark subsector.
With $\mu\ne 0$, the latter group is broken down to $\SU(2)_+\times \SU(2)_-\times U(1)$,
where $SU(2)_+$ has been described above, $SU(2)_-$ is the corresponding
symmetry between negative twist quarks $d_S$ and $u_V$, 
while the U(1) acts oppositely on positive
and negative twist quarks. We only make use of the $\SU(2)_+$ in the following.

We note in passing that one could use a formulation in
which all OS quarks, both up and down, are valence quarks.
This would lead to a clear separation between valence and sea quarks, whereas in the
formulation described above, $u_S$ is playing both roles.
The disadvantage of this  ``fully valence'' approach is that it would require more fields,
since one would need to introduce
{\em two} valence doublets and their corresponding ghosts,
leading to the field $Q$ having 10 components. 
We prefer to work with the minimal formulation given in Eq.~(\ref{eq:Q}).

The effective chiral low-energy theory describing the interactions of the pseudo-Goldstone
bosons and fermions (PGBs and PGFs) in the present context has been developed in several works.
The construction of PQ\chpt\ in the continuum was worked out in Ref.~\cite{Bernard:1993sv}; 
see Refs.~\cite{Sharpe:2006pu,Golterman:2009kw} for reviews. 
The systematic extension of \chpt\ to include the effects of discretization errors
for TM fermions was made in Refs.~\cite{Munster:2003ba,Scorzato:2004da,Sharpe:2004ps,Walker-Loud:2005ymg}, 
and applied to pion scattering in  Ref.~\cite{Buchoff:2008hh}. 
The combination of partial quenching and twisted-mass effects was
described in Ref.~\cite{\HSconstraint} and extended in Ref.~\cite{Walker-Loud:2009fuz}.
Here we simply quote the resulting form of the effective theory.

Before doing so we comment on the appropriate power counting needed to develop TM\chpt.
As noted in the introduction, when using near-physical quark masses, one has
$m_q \sim a^2$, where here we are leaving factors of $\Lambda_{\rm QCD}$ implicit.
Combined with the usual $p$-regime power counting of \chpt, this leads to 
\begin{align}
\begin{split}
\text{Leading order (LO):}\ & m_q \sim p^2 \sim a^2\,,
\\
\text{Next-to-leading order (NLO):}\ & m_q a \sim p^2 a \sim a^3\,,
\\
\text{Next-to-next-to-leading order (NNLO):}\ & m_q^2 \sim m_q p^2  \sim p^4
\sim m_q a^2 \sim p^2 a^2 \sim a^4\,.
\end{split}
\label{eq:powercounting}
\end{align}
The regime in which this power counting holds is referred to as the LCE 
(large cutoff effects) or ``Aoki'' regime.

For the most part we shall need only the LO chiral Lagrangian. 
This is given, in Euclidean space, by~\cite{BRS03,SS,\HSconstraint} 
\begin{equation}
\begin{split}
\cL_{\rm LO} &= \frac{f^2}4 {\rm str}\!\left(\partial_\mu \Sigma^\dagger \partial_\mu \Sigma \right)
-\frac{f^2}{4} {\rm str}\!\left(\chi \Sigma^\dagger+\Sigma \chi^\dagger\right) 
+ \cV_{a^2} \,,
\\
\cV_{a^2} &=  - \hat a^2 W_6' \,{\rm str}\!\left(\Sigma+\Sigma^\dagger\right)^2
- \hat a^2 W_7'\, {\rm str}\!\left(\Sigma-\Sigma^\dagger\right)^2
- \hat a^2 W_8'\, {\rm str}\!\left(\Sigma^2+\Sigma^{\dagger2}\right)\,,
\end{split}
\label{eq:LchptLO}
\end{equation}
where $\Sigma \in \SU(4|2)$ is the field that contains PGBs and PGFs,
 ``str'' denotes supertrace (or ``strace'' for short),
$\chi = 2 B_0 M$, with $M$ the renormalized mass matrix, 
and $f \approx 90\;$MeV and $B_0$ are the usual continuum LO low-energy coefficients (LECs),
while $\hat a= 2 W_0 a$,
with $W_0$, along with $W_6'$, $W_7'$, and $W_8'$, 
being LECs associated with discretization errors.
If we restrict ourselves to the unquenched subsector,
the only combination of the latter three LECs that appears
is $2W_6'+W_8'$. 

The form (\ref{eq:LchptLO}) holds for any partially quenched, 
$\cO(a)$ improved, Wilson-like LQCD action.
To obtain PQTM fermions, one uses the diagonal mass matrix
\begin{equation}
M = m_q e^{i\omega_m \tau_3^{VS}}\!,\quad
\tau_3^{VS} = {\rm diag}(\tau_3,\tau_3,\tau_3) = {\rm diag}(1,-1,1,-1,1,-1)\,,
\label{eq:MPQTM}
\end{equation}
where $m_q$ is the physical quark mass, and $\omega_m$ is the input twist angle.
These are related to the bare parameters $m_0$ and $\mu_0$ in the standard manner 
(see, e.g., Ref.~\cite{Sharpe:2004ps});
all we need here is that for maximal twist, $\omega_m=\pi/2$, 
we have $m_q= Z_P^{-1} \mu_0/a$.

The vacuum is determined at LO by minimizing the potential, including both the mass term
and $\cV_{a^2}$. This leads to
\begin{equation}
\Sigma_0 \equiv \langle 0|\Sigma|0\rangle
= \exp\left(i \omega_0 \tau_3^{VS} \right)\,,
\label{eq:Sigma0}
\end{equation}
with $\omega_0$ the vacuum twist angle.
In general $\omega_0$ differs from $\omega_m$ by $\cO(1)$, due to the competition
between the mass and $a^2$ terms in the potential, but for maximal twist one has
$\omega_0=\omega_m=\pi/2$. This is the case that we study henceforth.
There are corrections to this result at higher order in \chpt~\cite{Sharpe:2004ny}, 
but we will not need these in the following.

This discussion has ignored the presence of the strace in the terms in the potential.
Indeed, while the choice (\ref{eq:Sigma0}) minimizes the potential in the quark sector, 
it maximizes it in the ghost sector. Nevertheless,
as argued in Refs.~\cite{Sharpe:2001fh,Golterman:2005ie},
it is legitimate to proceed naively and develop perturbation theory 
by expanding about the extremum.
The arguments for this also show that, for the purposes of developing perturbation theory,
one can express $\cL_{\rm LO}$ in terms of $\Sigma$ and $\Sigma^\dagger$,
rather than the choice $\Sigma$ and $\Sigma^{-1}$ that is required for
a nonperturbative analysis~\cite{jac1,jac2}.

We next expand $\Sigma$ about its vacuum value to define the PG fields.
This is conveniently done using
\begin{equation}
\Sigma = \xi_0 \Sigma_{\rm ph} \xi_0\,,\quad
\xi_0 = \exp\left(i \omega_0 \tau_3^{VS}/2 \right)\,,
\quad
\Sigma_0=\xi_0^2\,,
\end{equation}
for if we then expand $\Sigma_{\rm ph}$ as
\begin{equation}
\Sigma_{\rm ph} = \exp( \sqrt2 i\Pi/f)\,,
\end{equation}
the fields in the PG matrix $\Pi$ have their standard flavor interpretation.
For example, the upper left $2\times2$ block contains the unquenched pion field,
\begin{equation}
\pi_{\rm SS} = \begin{pmatrix} 
\tfrac1{\sqrt2}\pi^0_{\rm SS} & \pi^+_{\rm SS}\\ \pi^-_{\rm SS}=(\pi^+_{\rm SS})^\dagger & -\tfrac1{\sqrt2}\pi^0_{\rm SS} 
\end{pmatrix}\,,
\label{eq:piSS}
\end{equation}
as well as a singlet component. This was the basis used in Ref.~\cite{Hansen:2011kk}.
We shall instead use a different basis for the neutral fields
(and different nomenclature for the charged fields).
Since $\Pi$ is straceless, there are five linearly independent neutral fields.
We choose these to be $\pi_{++}^{0}$ and $\pi_{--}^{0}$,
which are the neutral components of the OS pion isotriplets composed of quarks of definite twist,
together with three neutral fields composed of states with differing twist:
$\eta_{4}$, which is an SU(4) generalization of the $\eta$;
$\phi_{0}$, which is a superposition of the two ghost-antighost pairs;
and $\phi_{1}$, which spans the quark and ghost sectors.
\begin{equation}
\Pi = 
 \begin{pmatrix}
\frac{\pi^0_{++}}{\sqrt2}\!+\!\frac{\eta_4}2 \!+\!\frac{\phi_1}2
 & \pi^+_{-+} & \pi^+_{++} & \pi^{uu}_{-+}  & \omega_{15} & \omega_{16} \\
(\pi^+_{-+})^\dagger %\pi^-_{+-} 
& -\frac{\pi^0_{--}}{\sqrt2}\!-\!\frac{\eta_4}2 \!+\!\frac{\phi_1}2
 & (\pi^{dd}_{-+})^\dagger %\pi^{dd}_{+-} 
 & (\pi^+_{--})^\dagger % \pi^-_{--} 
 & \omega_{25} & \omega_{26} \\
(\pi^+_{++})^\dagger %\pi^-_{++} 
& \pi^{dd}_{-+} &-\frac{\pi^0_{++}}{\sqrt2}\!+\!\frac{\eta_4}2 \!+\!\frac{\phi_1}2
 & (\pi^+_{+-})^\dagger & \omega_{35} & \omega_{36} \\
(\pi^{uu}_{-+})^\dagger %\pi^{uu}_{+-} 
& \pi^+_{--} & \pi^+_{+-} 
& \frac{\pi^0_{--}}{\sqrt2}\!-\!\frac{\eta_4}2 \!+\!\frac{\phi_1}2
 & \omega_{45} & \omega_{46} \\
\omega_{51} & \omega_{52} &\omega_{53} &\omega_{54} & \frac{\phi_0}{\sqrt2} \!+\! \phi_1 
 &\phi_{56} \\
\omega_{61} & \omega_{62} &\omega_{63} &\omega_{64} & \phi_{56}^\dagger % \phi_{65} 
 & -\frac{\phi_0}{\sqrt2} \!+\! \phi_1
\end{pmatrix}\,.
\label{eq:Pi}
\end{equation}
The subscripts on the various pion fields indicate the twists of the component 
antiquark and quark fields, respectively.
This matrix is straceless and hermitian, with the $\omega_{jk}$ being PGFs,
while the $\phi$ fields are PGBs that have propagators with an unphysical overall sign.
We stress that $\pi^+_{\rm SS}$ and $\pi^+_{-+}$ are different notations for the same field,
as can be seen by comparing Eqs.~(\ref{eq:piSS}) and (\ref{eq:Pi}).
The choice of which notation we use in the following 
depends on which feature of the results that we wish to emphasize.

We have assumed in the above that we can treat $\Pi$ as a hermitian field, so that
$\Sigma_{\rm ph}^\dagger=\exp(-\sqrt2 i \Pi/f)$.
For the bosonic fields in $\Pi$ this is standard,
while in the fermionic sector it is a convenient convention.
Specifically we are defining hermitian conjugation such that
$\omega_{jk}=\omega_{kj}^\dagger$. This is a convention since
$\omega_{jk}$ and $\omega_{kj}$ are treated as independent Grassman fields
in Euclidean functional integrals.

The lattice simulations of Ref.~\cite{Liu:2016cba}
use the pion fields $\pi^+_{++}$, $\pi^0_{++}$ and $\pi^-_{++}$, and it is on these
that we mainly focus below. Due to a discrete symmetry, we could, however, have
equally well used the $\pi^a_{--}$ states, i.e. those composed of fields with negative
twist.

Rewriting the Lagrangian in terms of $\Sigma_{\rm ph}$ leads, at maximal twist, to
\begin{multline}
\cL_{\rm LO} = \frac{f^2}4 {\rm str}\!
\left(\partial_\mu \Sigma_{\rm ph}^\dagger \partial_\mu \Sigma_{\rm ph} \right)
-\frac{f^2 B_0 m_q}{2} {\rm str}\!\left(\Sigma_{\rm ph}^\dagger+\Sigma_{\rm ph}\right) 
- \hat a^2 W_6'  \left[{\rm str}\!
\left(\Sigma_0\Sigma_{\rm ph}+\Sigma_{\rm ph}^\dagger\Sigma_0^\dagger\right)\right]^2
\\
- \hat a^2 W_7'\left[ {\rm str}\!\left(\Sigma_0\Sigma_{\rm ph}-
\Sigma_{\rm ph}^\dagger\Sigma_0^\dagger\right)\right]^2
- \hat a^2 W_8'\, {\rm str}\!\left(\Sigma_0\Sigma_{\rm ph}\Sigma_0\Sigma_{\rm ph}
+\Sigma_{\rm ph}^{\dagger}\Sigma_0^\dagger\Sigma_{\rm ph}^\dagger\Sigma_0^\dagger\right)\,,
\label{eq:chlag}
\end{multline}
Expanding in powers of $\Pi$, the quadratic term is
\begin{equation}
\cL_{\rm LO} \supset \tfrac12 {\rm str}\!\left(\partial_\mu \Pi \partial_\mu \Pi\right)
+ B_0 m_q {\rm str}\! \left( \Pi^2 \right)
- \tfrac12 {w_6' f^2}\left[ {\rm str}\!\left(\tau_3^{VS} \Pi\right)\right]^2
- \tfrac14 {w_8' f^2 }\left[ {\rm str}\!\left(\tau_3^{VS} \Pi \tau_3^{VS} \Pi\right)
+ {\rm str}\left( \Pi^2\right) \right]\,,
\label{eq:Lquad}
\end{equation}
while the quartic term is
\begin{multline}
\cL_{\rm LO} \supset 
\frac1{6 f^2} \str\left[ (\partial_\mu \Pi) \Pi (\partial_\mu \Pi) \Pi - (\partial_\mu \Pi)^2 \Pi^2 \right]
- \frac{B_0 m_q}{6 f^2} \str\left(\Pi^4\right)
+ \frac{w_6'}3 \str \left(\tau_3^{VS} \Pi\right) \str\left(\tau_3^{VS} \Pi^3\right)
\\
+ \frac{w'_7}4 \str\left(\tau_3^{VS} \Pi^2\right) \str\left(\tau_3^{VS} \Pi^2\right)
+ \frac{w'_8}{24} \left[
3 \str\left(\tau_3^{VS} \Pi^2 \tau_3^{VS} \Pi^2\right)
+ 4 \str\left(\tau_3^{VS} \Pi \tau_3^{VS} \Pi^3\right)
+  \str\left(\Pi^4\right)
\right]\,.
\label{eq:Lquartic}
\end{multline}
Here we have converted to the dimensionless LECs
\begin{equation}
w'_j = \frac{16 \hat a^2 W'_j}{f^4} \qquad (j=6,7,8)\,.
\end{equation}
The cubic term vanishes at maximal twist, a point explained in Appendix~\ref{sec:NLO}.

\begin{comment}
The relation of the pion fields in \chpt\ to operators in the underlying theory can be
worked out using the spurion method. The resulting correspondence
is given for $j,k=1\!-\!4$ by (see, e.g. Ref.~\cite{Hansen:2011mc})
$\bar q_j \gamma_5 q_k(x) = c (\Sigma-\Sigma^\dagger)_{kj}$, 
with $c$ a known constant that will cancel in the ratios considered below.
Choosing $j\ne k$ for simplicity, the right-hand side of this relation is proportional to $\pi_{kj}$,
up to chiral corrections proportional to $\pi^2/f^2$.
As discussed in Sec.~C.1 of Ref.~\cite{Hansen:2011mc}, these correction terms lead
to subleading contributions to the correlation functions that lie beyond the accuracy that we
consider. Thus we effectively have the correspondence $\pi_{kj} \sim \bar q_j\gamma_5 q_k$,
and, similarly, $\pi^0 \sim \tfrac1{\sqrt2}(\bar q_1 \gamma_5 q_1 -\bar q_2 \gamma_5 q_2)$
and $\eta_4 \sim\tfrac12 (\bar q_1 \gamma_5 q_1 + \bar q_2 \gamma_5 q_2 - \bar q_3\gamma_5 q_3
-\bar q_4 \gamma_5 q_4)$, with a common constant of proportionality.
Using this correspondence, the definitions given below apply both in the
underlying theory, PQQCD, and in the effective theory, PQ\chpt, and we will move 
between these two representations as needed.
\end{comment}

\section{Leading-order TMPQ\chpt\ results}
\label{sec:LO}

In this section we use the chiral Lagrangian given in Eq.~(\ref{eq:chlag}) to determine the LO results
for propagators, masses and scattering amplitudes.
We also briefly discuss the mixing of two OS pions having $I=0$ with a single unquenched
neutral pion, which requires working at NLO in \chpt.

\subsection{Propagators and masses}

We first read off the propagators and particle masses from Eq.~(\ref{eq:Lquad}).
For fields labeled as lower-case pions in Eq.~(\ref{eq:Pi}), 
the propagators take the standard form $1/(p^2+M^2)$, with the masses being
\begin{align}
M(\pi^\pm_{-+})^2 &= M(\pi^{uu}_{-+})^2 =M(\pi^{dd}_{-+})^2 = M(\pi^\pm_{\rm SS})^2 = 2 B_0 m_q \equiv {(M_{\rm SS}^{\pm})}^2 \,,
\label{eq:MpipmSS}
\\
M(\pi^\pm_{++})^2 &= M(\pi^0_{++})^2= M(\pi^\pm_{--})^2=M(\pi^0_{--})^2 = 
2 B_0 m_q  - w_8' f^2 \equiv M_{\rm OS}^2\,,
\label{eq:Mpipmpp}
%\\
%M(\eta_4)^2 &=  2 B_0 m_q - (4 w_6' + w_8')f^2\,.
\end{align}
As noted in the first line, the masses of all the pions that are
composed of quarks and antiquarks with opposite twists coincide,
and contain no discretization errors at LO.
This includes the charged sea-quark pions.
These masses differ by $\cO(a^2)$ from those of the pions
in which the quarks and antiquarks have the same twist, i.e. the results in the second line.
Such pions are collectively called OS pions in Ref.~\cite{Liu:2016cba},
and thus we have introduced the common nomenclature $M_{\rm OS}$
(adapted from that in Ref.~\cite{Liu:2016cba}, where $M_\pi^{\rm OS}$ is used).

As an aside, we note that $M_{\rm OS}$ is also the mass obtained from
the pole in the correlator of the unquenched neutral pion, $\pi_{\rm SS}^0$,
{\em if one keeps only the connected Wick contraction}.
This mass is denoted $m_\pi^{0,\rm conn}$ in Ref.~\cite{\HSconstraint}.
The separation of contractions can be formulated in a partially quenched theory,
and one finds
$m_\pi^{0,\rm conn}=M_{\rm OS}$~\cite{\HSconstraint}.

A prediction from Eqs.~(\ref{eq:MpipmSS}) and (\ref{eq:Mpipmpp}) is that
\begin{equation}
\Delta_{\rm OS} = M_{\rm OS}^2 - {(M_{\rm SS}^{\pm})}^2 = - w_8' f^2 \,,
\label{eq:Mpisplit}
\end{equation}
i.e. that there is a mass-independent $\cO(a^2)$ offset between 
the charged sea-quark and OS pion squared masses.
Since $w_8'$ is known to be negative~\cite{jac1,jac2,\HSconstraint}, this offset is positive.
To our knowledge, the accuracy of this prediction in practice has not been discussed previously.
It can be tested using the results quoted in Table II of Ref.~\cite{Liu:2016cba}, and
this is done in Table~\ref{tab:Mpisplit}.
Since the two masses in the difference are certainly correlated, we cannot determine 
the error on the difference. Nevertheless, it is clear that, to high precision, the offset is constant.
This indicates that higher-order corrections to Eq.~(\ref{eq:Mpisplit}),
which, among other effects, would lead to a linear dependence on $\mu_0$, are small.
The sign of the offset is as predicted, and its value is such that the $m_q$ and $a^2$ contributions
to $M_{\rm OS}^2$ are comparable for physical quark masses, 
as already noted in the introduction.
Numerically, we find $\sqrt{\Delta_{\rm OS}}/a \approx 200\;$MeV, which is 
comparable to $\Lambda_{\rm QCD}$ and thus of the expected size.

\begin{table}[htp]
\begin{center}
\begin{tabular}{cccc}
$\mu_0$ &  $a M(\pi^\pm_{\rm SS})$ & $a M_{\rm OS}$ 
& $a^2 \Delta_{\rm OS}$ \\
\hline
0.0009 & 0.06212(6) & 0.11985(15) & 0.0105
\\
0.003   & 0.11197(7)   & 0.15214(11) & 0.0106
\\
0.006 & 0.15781(15) & 0.18844(24)  & 0.0106
\end{tabular}
\end{center}
\caption{Numerical results for the mass splitting defined in Eq.~(\ref{eq:Mpisplit}).
These are given in the final column, and determined from the results in the
second and third columns, which are taken from Ref.~\cite{Liu:2016cba}.
The first column gives the bare twisted mass, with the top row corresponding
approximately to physical-mass charged pions. The lattice spacing is
$a=0.931(8)\;$fm~\cite{ETM:2015ned} }
\label{tab:Mpisplit}
\end{table}%

The results for the PGF masses match those of the charged pions, and, in particular,
depend only on the relative sign of the twist of the component quark and ghost-quark,
\begin{align}
M(\omega_{16})^2&=M(\omega_{25})^2 = M(\omega_{36})^2=M(\omega_{45})^2
= 2B_0 m_q\,,
\\
M(\omega_{15})^2&=M(\omega_{26})^2 = M(\omega_{35})^2=M(\omega_{46})^2
=2 B_0 m_q  - w_8' f^2\,.
\end{align}
The propagator has the bosonic form, with the overall sign depending on the ordering
of $\omega_{jk}$ and $\omega_{kj}$.
Specifically, when ordered with $\omega_{jk}$ preceding $\omega_{kj}$,
an overall minus sign arises if $j<k$.

The ghost-ghost PGBs $\phi_1$ and $\phi_{56}$ have propagators with unphysical
overall signs, $-1/(p^2+M^2)$, due to the supertrace. Their masses are
\begin{equation}
M(\phi_{56})^2 = 2 B_0 m_q\,, \qquad
M(\phi_1)^2 = 2 B_0 m_q - w_8 f^2 \,,
\end{equation}
again matching the pattern for the $\pi$ fields.

Finally, we consider the $\eta_4$ and $\phi_0$. Here there is mixing, due to the
$w_6'$ term in Eq.~(\ref{eq:Lquad}). 
In the basis $\{\eta_4,\phi_0\}$, the propagator is given by the inverse of
\begin{equation}
\begin{pmatrix} p^2 + 2B_0m_q-w'_8 f^2 & 0 \\
                        0 &  -(p^2 + 2B_0m_q-w'_8 f^2) \end{pmatrix}
-w'_6 f^2 \begin{pmatrix} 4 & - 2\sqrt2 \\ -2 \sqrt2 &2 \end{pmatrix}\,,
\label{eq:mixing}
\end{equation}
which is
\begin{equation}
 \frac1{p^2 + M(\pi^0_{\rm SS})^2} \begin{pmatrix} 1 & 0 \\ 0 & -1  \end{pmatrix}
+\frac{2 w'_6 f^2}{[p^2 + M(\pi^0_{\rm OS})^2][p^2 + M(\pi^0_{\rm SS})^2] }
\begin{pmatrix} 1 & \sqrt2 \\ \sqrt2 &2 \end{pmatrix}\,.
\label{eq:mixingprop}
\end{equation}
Here
\begin{equation}
M(\pi_{\rm SS}^0)^2 = 2 B_0 m_q - (2 w_6'+w_8') f^2 \equiv (M_{\rm SS}^{0})^{2}
\label{eq:Mpi0SS}
\end{equation}
is the mass of the neutral unquenched pion.
The appearance of this mass can be understood as due to hairpin diagrams with
loops of sea quarks summed to all orders.
We also observe that the second term in the propagator,
Eq.~(\ref{eq:mixingprop}), involves the double poles characteristic of PQ\chpt.

A nontrivial check on the results just presented is obtained by directly
calculating the propagator of $\pi^0_{\rm SS}$, which can be done using the relation
[obtained by comparing the representations of the upper left block of $\Pi$
given by Eqs.~(\ref{eq:piSS}) and (\ref{eq:Pi})]
\begin{equation}
\pi^0_{\rm SS} = \tfrac12 \pi^0_{++} + \tfrac12 \pi^0_{--} + \tfrac1{\sqrt2} \eta_4\,.
\end{equation}
Using the results above, this propagator is given by
\begin{equation}
2 \times \frac14 \frac1{p^2 + (M_{\rm OS}^{0})^{2}}
+ \frac12 \left\{ \frac1{p^2 + (M_{\rm SS}^{0})^{2}}
+\frac{2 w'_6 f^2}{[p^2 + (M_{\rm OS}^{0})^{2}][p^2 + (M_{\rm SS}^{0})^{2}] } \right\}
= \frac1{p^2 + (M_{\rm SS}^{0})^{2}}\,,
\end{equation}
which is the desired result.

We close this subsection by noting that, by using a clover term in the fermion action,
the breaking of the physical isospin symmetry in the unquenched theory is greatly reduced,
with $M_{\rm SS}^{\pm} \approx M_{\rm SS}^0$ (see Table II of Ref.~\cite{Liu:2016cba}).
Thus the combination $2 w'_6+w'_8$, which controls isospin breaking in the
physical sector, almost vanishes. We stress, however, that this does not imply
that discretization errors in all quantities are similarly small, since, as we have seen 
from Table~\ref{tab:Mpisplit}, $w'_8$ itself is not small.

\subsection{Scattering amplitudes}
\label{sec:LOamp}

We next consider LO scattering amplitudes, which
can be determined from the quartic term in $\cL_{\rm LO}$, given in Eq.~(\ref{eq:Lquartic}).
Results for scattering amplitudes in the unquenched sector have already been
given in Ref.~\cite{Buchoff:2008hh}, and we have checked that we agree with these.
What we focus on here are the scattering amplitudes of OS pions,
and in particular those composed of quarks with positive twist.
These are the amplitudes studied numerically in Ref.~\cite{Liu:2016cba},
and we wish to determine the form of the $\cO(a^2)$ contributions to them.

For the scattering of OS pions, the nonderivative terms in $\cL_{\rm LO}$
collapse down to a contribution that is of the same form as that in the continuum,
\begin{equation}
c_{\rm OS} \left[ \frac{(\pi^0_{++})^2}2 + \pi^+_{++} \pi^-_{++} \right]^2
\,,
\label{eq:cOSterm}
\end{equation}
except that the overall coefficient,
\begin{equation}
c_{\rm OS} 
=
-\frac{2 B_0 m_q}{6 f^2} + w'_7 + \tfrac23 w'_8
=
-\frac{M_{\rm OS}^2}{6 f^2} + w'_7 + \tfrac12 w'_8
\,,
\label{eq:cOSres}
\end{equation}
includes $a^2$ corrections.
As expected, the form in Eq.~(\ref{eq:cOSterm}) is invariant under the OS isospin group, 
${\rm SU}(2)_+$, so that there is no mixing between $I=0$ and $I=2, I_z=0$ amplitudes.
On the other hand, these results imply that
 there is an $\cO(a^2)$ contribution to the scattering of two OS $\pi^+$ mesons,
unlike for the corresponding sea pions.

Including the kinetic term, the full LO scattering amplitudes for the different choices
of OS isospin are given by
\begin{align}
\cA_{I=2}^{\rm LO} 
&=  \frac1{f^2} (\underbrace{-s + \tfrac43 M_{\rm OS}^2)}_{\rm kinetic\ term} - 4 c_{\rm OS}\,,
\\
&= \frac1{f^2} ( - s + 2 M_{\rm OS}^2) - 4 w'_7  - 2 w'_8 \,,
\\
&= \frac1{f^2} \left[ - s + 2 M_{\rm OS}^2 + 2 \Delta_{\rm OS} -  4 f^2 w'_7 \right]  \,,
\label{eq:ALOI2}
\\
\cA_{I=1}^{\rm LO} &= \frac{t-u}{f^2} 
=  - (p_1-p_2) \cdot (k_1-k_2)
\label{eq:ALOI1}
\\
\cA_{I=0}^{\rm LO} 
&=  \frac1{f^2} ( 2 s -  M_{\rm OS}^2) - 10 w'_7 - 5 w'_8\,,
\\
&=  \frac1{f^2} \left[2 s -  M_{\rm OS}^2 + 5 \Delta_{\rm OS} - 10 f^2 w'_7 \right] \,.
\label{eq:ALOI0}
\end{align}
The $I=1$ result comes only from the kinetic term, since it requires momentum dependence, and
thus is the same as in the continuum.
The $I=2$ and $I=0$ results contain discretization errors from both $w_7'$ and $w_8'$,
although the former can be rewritten in terms of the measured mass difference $\Delta_{\rm OS}$,
as shown by the final forms for each quantity.
Also useful are the $s$-wave scattering lengths, 
which are proportional to the amplitudes at threshold, where $s=4 M_{\rm OS}^2$.
These are given by
\begin{comment}
\begin{align}
\cA_{I=2}^{\rm LO, th} 
&= -2\frac{ M_{\rm OS}^2}{f^2}  + 2 \frac{ \Delta_{\rm OS}}{f^2}  - 4 w'_7 \,,
\\
\cA_{I=0}^{\rm LO, th} 
&= 7 \frac{M_{\rm OS}^2}{f^2}
+ 5 \frac{ \Delta_{\rm OS}}{f^2} - 10 w'_7 \,.
\end{align}
\end{comment}
\begin{align}
M_{\rm OS} \; a_{0, I=2}^{\rm LO}
&= \frac{ M_{\rm OS}^2-\Delta_{\rm OS}+ 2 f^2 w'_7 }{16\pi f^2} 
\label{eq:a0I2}
\\
M_{\rm OS} \; a_{0, I=0}^{\rm LO}
&= - \frac{7M_{\rm OS}^2+5\Delta_{\rm OS}- 10 f^2 w'_7 }{32 \pi f^2} \,.
\label{eq:a0I0}
\end{align}
Here we are using the sign convention, standard in most of the LQCD literature
(although not followed in Ref.~\cite{Liu:2016cba}),
that a positive scattering length corresponds to a repulsive interaction.

We draw two conclusions from these results.
First, the $I=0$ OS amplitude does contain $\cO(a^2)$ terms.
Since these involve both $\Delta_{\rm OS}$ (which is known) and $w'_7$ (which is not),
we do not know \emph{a priori} the size of these corrections.
They could be comparable to the physical $M_{\rm OS}^2$ term, as for $M_{\rm OS}$ itself,
or they could be smaller, as in the difference $M_{\rm SS}^{\pm} - M_{\rm SS}^{0}$.
It is not possible to determine which option holds from the results of Ref.~\cite{Liu:2016cba} alone,
since they are available only for two quark masses, and at a single lattice spacing,
and have relatively large errors.

Our second conclusion is that the size of the discretization errors for $I=0$ scattering 
can be determined by calculating the $I=2$ amplitude for OS pions. 
In particular, if one calculates $a_{0, I=2}$ and
compares to the LO prediction (\ref{eq:a0I2}), then one can determine $w'_7$ (given
the result obtained above for $\Delta_{\rm OS}$), and use this to remove the 
$\cO(a^2)$ errors from $a_{0, I=0}$, Eq.~(\ref{eq:a0I0}).
A more direct way of proceeding is to make use of the fact that the $\cO(a^2)$ terms in the
two amplitudes are proportional, so that LO discretization errors
cancel in the following linear combinations:
\begin{align}
\cA_{I=0}^{\rm LO} - \tfrac52 \cA_{I=2}^{\rm LO} &=  \frac{\tfrac92 s - 6 M_{\rm OS}^2}{f^2}\,,
\\
M_{\rm OS} \left( a_{0, I=0}^{\rm LO} - \tfrac52 a_{0, I=2}^{\rm LO}\right) &=
- \frac{3 M_{\rm OS}^2}{8 \pi f^2}\,.
\end{align}

To our knowledge a calculation of $a_{0, I=2}$ using OS pions has not been carried out;
indeed, there has previously been no motivation to do so given that $I=2$ scattering can be studied
with unquenched pions.
Such a calculation should, however, be straightforward, since it involves only a subset of the
Wick contractions needed for the $I=0$ calculation, and this subset involves only quark-connected
diagrams.

One might be concerned that higher-order corrections in \chpt\ could be significant.
In this regard, we have two comments. First, the results in Table~\ref{tab:Mpisplit}
provide partial evidence that the LO prediction is reliable, as there is no significant dependence
on $\mu_0$. This does not, however, rule out large $a^4$ corrections.
Second, there are no NLO contributions to the scattering amplitudes discussed in
this subsection. The reason for this is discussed in the following subsection.
The first corrections occur at NNLO, an example of which is discussed in Sec.~\ref{sec:loop}.

\subsection{Mixing with states with vacuum quantum numbers}
\label{sec:Mixing}

A significant challenge in the lattice calculation of Ref.~\cite{Liu:2016cba}
is the presence of mixing between the OS $I=0$ state and a single neutral unquenched pion
($\pi_{\rm SS}^0$ in our notation).
In the continuum, such mixing violates parity and G-parity, but with TM fermions,
lattice artifacts break parity and the physical isospin symmetry in such a way that mixing is allowed.
It does not occur, however, at LO in \chpt\ when working at maximal twist, as is evident
from the absence of cubic vertices in Sec.~\ref{sec:setup}.
A natural question, therefore, is at what order does this mixing occur.

The answer is that it occurs at NLO in the expansion described in Eq.~(\ref{eq:powercounting}).
To see this one must write down all allowed NLO terms in the Lagrangian and expand
in the field $\Pi$. There are many such terms, each with a different combinations of LECs, 
all of which correspond to artifacts of discretization and are unknown. 
Thus we do not think it is useful to explicitly calculate the mixing amplitude.
Instead, we determine the overall chiral scaling of the terms that contribute.

The form of the allowed NLO terms is given in Appendix~\ref{sec:NLO},
Eqs.~(\ref{eq:LNLOW}) and (\ref{eq:La3}).
If we expand these out we obtain linear terms of the form
\begin{equation}
(a^3 + a m_q){\rm str}\!(\Pi \tau_3^{VS}) \,
\label{eq:NLOlin}
\end{equation}
and the following cubic terms
\begin{multline}
a^3 {\rm str}\! (\Pi \tau_3^{VS})^3\,, \ \ 
a^3 {\rm str}\! (\Pi \tau_3^{VS}\Pi \tau_3^{VS}\Pi \tau_3^{VS})\,, \ \ 
a^3 {\rm str}\! (\Pi \tau_3^{VS}\Pi \tau_3^{VS}) {\rm str}\! (\Pi \tau_3^{VS})\,, \ \ 
(a^3 + a m_q) {\rm str} \! (\Pi^2) {\rm str}(\Pi \tau_3^{VS})\,, \ \ 
\\
(a^3 + a m_q) {\rm str} \! (\Pi) {\rm str}(\Pi^2 \tau_3^{VS})\,, \ \
(a^3 + a m_q)  {\rm str}\! (\Pi^3 \tau_3^{VS})\,, \ \ 
a  \;{\rm str}\! (\partial \Pi  \partial \Pi (\Pi \tau_3^{VS} + \tau_3^{VS} \Pi))\,, \ \ 
a  \;{\rm str}\! (\partial \Pi  \partial \Pi ){\rm str}\!( \Pi \tau_3^{VS})\,. \ \ 
\end{multline}
As is discussed in the appendix, there are no terms with even powers of $\Pi$.
In these equations we are being schematic, showing simply the overall structures that appear
and the form of the coefficients.
Ignoring for now the linear term, we see that the three-pion vertices appear for nonzero $a$,
and it is straightforward to see that they lead to vertices of the forms
$\pi_{++}^+ \pi_{++}^- \pi^0_{\rm SS}$ and $\pi_{++}^0 \pi_{++}^0 \pi^0_{\rm SS}$.
Thus we find that the mixing discussed in Ref.~\cite{Liu:2016cba} occurs at NLO,
and that the on shell mixing amplitude is 
proportional to $a^3$ and $a M^2$, where $M$ is a generic meson mass.

Now we return to the linear term, Eq.~(\ref{eq:NLOlin}),
and argue that it leads to contributions scaling in the same fashion as those
from the cubic terms just discussed.
This leads to a mixing of $\pi^0_{\rm SS}$ with
the vacuum, and thus to a shift in the twist angle of $\cO(a)$.
%proportional to $(a^3 + a M)/(a^2 + M^2)$.
This in turn leads to additional contributions involving odd powers of $\Pi$
that are fed down from the even-powered LO vertices. 
One way of seeing this is to take, say, a four point LO vertex
and absorb one of the pions into the vacuum, leading to an additional factor of
$a^3+aM^2$ from the absorption vertex and $1/(a^2+M^2)$ from the propagator.
Combining this with the $a^2+M^2$ from the initial vertex one ends up, crudely
speaking, with the same $a^3+aM^2$ scaling as that obtained from the cubic vertices.

\section{Applicability of two-particle quantization condition}
\label{sec:loop}

In this section, we discuss in which channels 
it is valid to use the two-particle quantization condition of L\"uscher
when using TM fermions with OS valence quarks.\footnote{%
The quantization condition breaks down above the first inelastic threshold,
where more than two particles can go on shell, and we consider here only the kinematic
regime in which this is not allowed.}
To do so, we apply the criteria that we recently developed in Ref.~\cite{\DSPQLuscher}.
Unitarity is an essential ingredient both in the derivation of the quantization condition,
and in the justification for the standard spectral representation of correlators.
The observation of Ref.~\cite{\DSPQLuscher} is that it is likely sufficient, however,
for unitarity to hold in the $s$ channel.
If the unitarity violation introduced by the use of a partially quenched theory
manifests itself only in the $t$ and $u$ channels, then this does not invalidate the
use of the quantization condition. 
In the case at hand, this question can be studied using \chpt.

The simplest criterion to implement of those proposed in Ref.~\cite{\DSPQLuscher} 
(which is the second of the criteria introduced in that work)
uses quark line diagrams, which trace the flavor indices in  \chpt\ calculations. 
The criterion states that if intermediate ghost quarks appear in the quark-line diagrams for
$s$-channel two-particle loops, then the quantization condition cannot be used.
The logic here is that ghost quarks act as a proxy for the presence of intermediate states that
differ from those on the external lines, indicating a breakdown of unitarity.

\begin{figure}
\includegraphics[width=\textwidth]{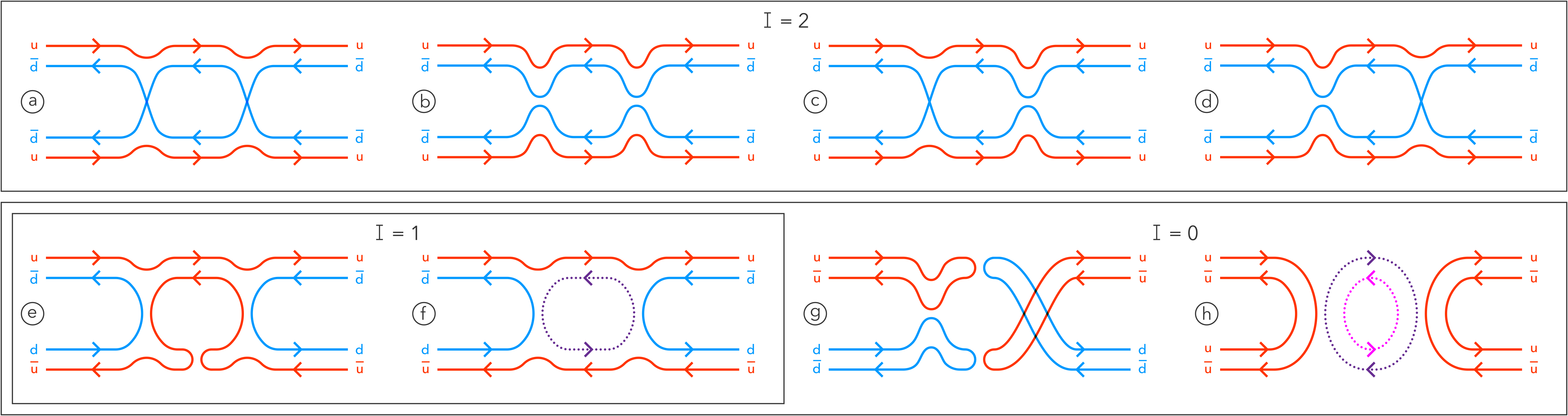}
\caption{Classes of topologically distinct $s$-channel one-loop quark-line diagrams 
contributing to  $I=2=I_z$, $I=1, I_z=0$, and $I=0$ scattering.
The choices of $I_z$ are for convenience.
All classes of diagram are shown for $I=2$ (up to exchange of the quark/antiquark flavors), while only a subset are shown for $I=1$ and $I=0$.
Labels and colors indicate quark or antiquark flavors, 
with lines tracing the (anti)quark flavors as they propagate through the diagram. 
Unlabeled dashed lines may have any flavor, and may represent sea, valence, or ghost (anti)quarks. The hairpin vertices in diagrams (e) and (g) implicitly contain an infinite sum of intermediate sea quark, valence and ghost loops.}
\label{fig:quarklines}
\end{figure}

In Fig.~\ref{fig:quarklines} we show examples of quark-line diagrams for the 
one-loop $s$-channel scattering process for OS pions having $I=2$, $1$ and $0$. 
For $I=2$, we show all the distinct classes of diagrams that appear, taking account of
the fact that the LO vertices, given in Eq.~(\ref{eq:Lquartic}), can have either one or two straces.
This leads to a greater number of quark-line diagrams than those appearing
in the continuum theory, where only single-trace vertices occur at LO. 
Nevertheless, none of the diagrams involve ghost quarks,
and thus using the quantization condition for $I=2$ OS pions is not ruled out.
This result extends to $s$-channel loops at any order in \chpt.

Next, we turn to $I=1$ and $I=0$, for which many more classes of diagram contribute. 
Since the presence of a single diagram involving ghost quarks is sufficient for our criterion
to invalidate the use of the quantization condition, we show in Fig.~\ref{fig:quarklines}
only examples of diagrams containing ghost loops. % for $I=1$ and $I=0$.
The ghost quarks are either explicit, as in diagrams (f) and (h),
or implicitly contained in the hairpin vertices, as in diagrams (e) and (g).
Since we have chosen to display diagrams for $I=1, I_z=0$,
each of the diagrams in this channel will also contribute to $I=0$ scattering.
There are many more types of diagram involving ghost quarks
that are not shown in Fig.~\ref{fig:quarklines}.
However, the subset of diagrams we do include is enough to 
conclude that the quantization condition {\em cannot} 
be used to study OS pion $I=1$ or $I=0$ scattering.

As noted in Ref.~\cite{\DSPQLuscher}, the diagrammatic criterion is not entirely foolproof, as the
contributions corresponding to a given unphysical quark-line diagram could cancel. 
However, in that work we also introduced two other criteria which do not suffer from such an ambiguity.
The first of these (which is the first of that work) 
restricts the form of the ratio of correlation functions given in Eq.~(\ref{eq:CorrRatio}),
and we do make use of it here.
Here we use the third criterion of Ref.~\cite{\DSPQLuscher}:
the two-particle quantization condition can only be valid if the PQ scattering amplitude satisfies
$s$-channel unitarity. We apply this criterion using one-loop results from \chpt.

\begin{figure}
\includegraphics[width=2 in]{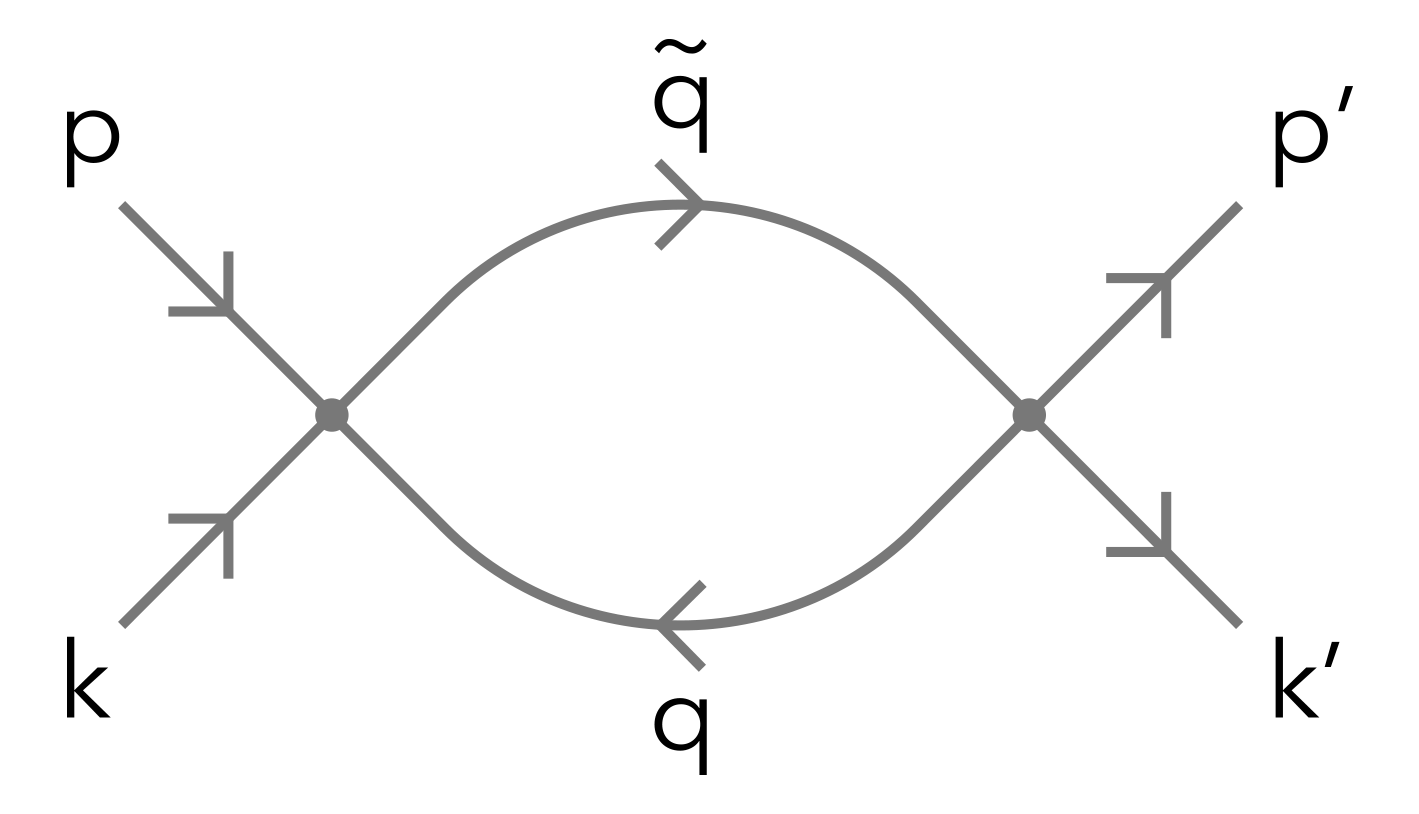}
\caption{One-loop $s$-channel scattering diagram with $p, k, p',$ and $k'$ labeling the external 4-momenta, and $q$ and $\tilde{q} = q + p + k$ labeling the internal 4-momenta.}
\label{fig:sloop}
\end{figure}

Since $s$-channel unitarity involves cutting diagrams only in the $s$ channel, it can be tested
at one-loop order simply by calculating the $s$-channel loop diagram, shown in Fig.~\ref{fig:sloop}.
The imaginary part of this loop must be proportional to the squared magnitude of the tree-level
amplitude, with the proportionality constant related to the phase space available at the chosen
value of $s$. One way in which unitarity can fail is if the particles 
in the loop have different
masses from those of the external particles, for then the phase space differs from that
appearing in the unitarity relation. 
%This provides a shortcut for checking unitarity in certain cases.

We begin with the $I=2$ channel, 
which we can pick out by studying $\pi^+_{\rm OS}\pi^+_{\rm OS}$ scattering.
The key point here is that only $\pi^+_{\rm OS}\pi^+_{\rm OS}$
intermediate states can appear in the $s$-channel loop.
This is clear from the diagrammatic quark-line analysis given above, 
%which traces the flavors labels of the fields in $\Pi$, 
but can be checked by direct calculation.
Specifically, we find that the contribution from the $s$-channel loop is 
\begin{equation}
\begin{split}
\cA^{(s1)}_{{\rm OS}, I=2} = & \frac1{f^4} \left( - s + 2 M_{\rm OS}^2 + 2 \Delta_{\rm OS} 
- 4 f^2 w'_7 \right)^2 J(s, M_{\rm OS}^{2}, M_{\rm OS}^{2}) 
\\
& 
+ \frac1{9 f^4} \left(5 s+24 w'_7 f^2 
- 12 M_{\rm OS}^2 - 12 \Delta_{\rm OS}\right) I\left(M_{\rm OS}^2\right),
\end{split}
\end{equation}
where $J(s, M_{\rm A}^{2}, M_{\rm B}^{2})$ is the $s$-channel loop integral
and $I\left(M_{\rm A}^{2}\right)$ is the tadpole integral that arises
when derivatives act on the internal legs. These are defined as follows,
\begin{align}
&J(s, M_{\rm A}^{2}, M_{\rm B}^{2}) \equiv \mu^\epsilon 
\int_{q} \frac{d^{4-\epsilon}q}{(2 \pi)^{4-\epsilon}} 
\frac{1}{(q^{2} + M_{\rm A}^{2})(\tilde{q}^{2} + M_{\rm B}^{2})}
\\
&I\left(M_{\rm A}^{2}\right) \equiv \mu^\epsilon 
\int_{q} \frac{d^{4-\epsilon}q}{(2 \pi)^{4-\epsilon}} 
\frac{1}{q^{2} + M_{\rm A}^{2}}.
\end{align}
Here, $q$ and $\tilde{q} = q + p + k$ are the momenta,
and $M_{\rm A}$ and $M_{\rm B}$ are the masses, that appear in the $s$-channel loop
(see Fig.~\ref{fig:sloop}).
Unitarity is studied by considering the coefficient of $J(s, M_{\rm OS}^{2}, M_{\rm OS}^{2})$,
since the tadpole function does not have an imaginary part.
Comparing to Eq.~(\ref{eq:ALOI2}), we see that the coefficient of $J$ is simply the square of
the LO scattering amplitude, $\cA_{I=2}^{\rm LO}$, 
and that the masses in the internal propagators are the same as those of the external legs.
Together this implies that $I=2$ scattering is consistent with $s$-channel unitarity.
Although this is only based on a one-loop calculation, 
we argue in Ref.~\cite{\DSPQLuscher} that it will hold to all orders.

Next we turn to the $I=1$ case. We have calculated the $s$-channel one-loop diagram for 
$\pi^+_{\rm OS} \pi^0_{\rm OS}$ scattering. The intermediate states that contribute are
$\pi^+_{\rm OS} \pi^0_{\rm OS}$, $\pi^+_{\rm SS} \pi^{dd}_{-+}$, $\pi^+_{\rm VV} \pi^{uu}_{+-}$,
$\omega_{15} \omega_{52}$, and $\omega_{16} \omega_{62}$.
A schematic form of the total result is sufficient for our purposes.
Let $\cA^{(s1)}_{\rm cont}(M_\pi)$ be the contribution 
from the loop in continuum QCD, which we know satisfies $s$-channel unitarity.
Then we find
\begin{align}
\cA^{(s1)}_{{\rm OS}, I=1} &= \tfrac12 \cA^{(s1)}_{\rm cont}(M_{\rm OS}) 
+ \tfrac12 \cA^{(s1)}_{\rm cont}(M_{\rm SS}^\pm)\,, \label{eq:IOneLoop}
\\
\cA^{(s1)}_{\rm cont}(M_\pi) & =
\frac{(t-u)}{6f^4} \left[\left(s-4 M_\pi^2 \right) 
J\left(s, M_\pi^2, M_\pi^2 \right) + 2 I\left( M_\pi^2 \right) \right]\,.
\end{align}
Since $M_{\rm SS}^\pm \ne M_{\rm OS}$, the phase-space for the second term of Eq.~(\ref{eq:IOneLoop})
does not match that required given that the external particles are OS pions, and $s$-channel unitarity fails.
This is consistent with the diagrammatic result from above.
We do stress, however, that unitarity
is recovered in the continuum limit, when $M_{\rm SS}^\pm = M_{\rm OS}$.

Finally, we consider the $I=0$ channel. 
Here the calculation is more involved, with many intermediate states.
We find that the contribution from $s$-channel loops is 
\begin{multline}
\cA^{(s1)}_{{\rm OS}, I=0} = \frac{3 s^2}{8 f^4} J\left(s, {(M_{\rm SS}^{\pm})}^2, {(M_{\rm SS}^{\pm})}^2 \right)
+ \frac{1}{8 f^4} J\left(s, M_{\rm OS}^2, M_{\rm OS}^2 \right) \Big(8 M_{\rm OS}^2 \left(8 f^2 w'_7 + 26 M_{\rm OS}^2 + 5 s\right)
\\
- 8 {(M_{\rm SS}^{\pm})}^2 \left(-8 f^2 w'_7 + 44 M_{\rm OS}^2 + 7 s\right) - 112 f^2 s w'_7 + 256 f^4 {w'_7}^2 + 160 {(M_{\rm SS}^{\pm})}^4 + 13 s^2\Big)
\\
- \frac{3}{f^4} \left(-3 {(M_{\rm SS}^{\pm})}^2 + 3 M_{\rm OS}^2 + {(M_{\rm SS}^{0})}^2 \right)^2 J\left(s, M_{\rm OS}^2, {(M_{\rm SS}^{0})}^2 \right)
\\
 + \frac{3}{2 f^4} \left(3 {(M_{\rm SS}^{\pm})}^2 - 2 \left(M_{\rm OS}^2 + {(M_{\rm SS}^{0})}^2 \right)\right)^2 J\left(s, {(M_{\rm SS}^{0})}^2, {(M_{\rm SS}^{0})}^2 \right)
\\
+ \frac{5 s}{12 f^4} I\left({(M_{\rm SS}^{\pm})}^2\right)
+ \frac{1}{36 f^4} I\left(M_{\rm OS}^2\right) \left(-336 f^2 w'_7 - 168 {(M_{\rm SS}^{\pm})}^2 + 120 M_{\rm OS}^2 + 65 s\right)\,.
\label{eq:IZeroLoop}
\end{multline}
It is straightforward to check that this form violates unitarity, both because the coefficients
of $J$ are not the square of the LO amplitude, Eq.~(\ref{eq:ALOI0}),
and because the masses in the arguments of $J$ vary from term to term.
Again, this is consistent with the diagrammatic analysis.

\section{Implications for previous work}
\label{sec:imp}

The conclusion from the previous section is that one cannot use the two-particle
quantization condition for $I=0$ OS pion scattering.
This would seem to invalidate the calculation of Ref.~\cite{Liu:2016cba},
and render the \chpt\ results of Sec.~\ref{sec:LO} irrelevant.
In this section we argue that the situation is not quite so dire.
The application of an invalid quantization condition introduces additional 
systematic errors into the results of Ref.~\cite{Liu:2016cba},
and these new errors can be approximately quantified.

The breakdown of $s$-channel unitarity leads to two distinct but related problems
if one nevertheless applies L\"uscher’s formalism.
The first concerns the fit that is used to extract the finite-volume energies.
In particular, if $\cO_{I}^{\dagger}$ is an operator that creates the pair of OS pions
with OS isospin $I$, and $\cO_{I}$ is the corresponding annihilation operator,
then the energies are extracted from
\begin{equation}
\langle \cO_{I}(\tau) \cO_{I}^{\dagger}(0) \rangle = \sum_{n=0}^{\infty} c_n e^{-E_n \tau}\,,
\end{equation}
where $E_0 \leq E_1 \leq E_2 \leq ... $, %are the energies in the two-particle spectrum
and the $c_n$'s are real and positive coefficients.
For physical theories, one can extract quantities of interest such as the energy shift
$\delta E_0 = E_0 - 2M_\pi$ using a ratio of correlation functions
\begin{equation}
R(\tau) = \frac{\langle \cO_{I}(\tau) \cO_{I}^{\dagger}(0) \rangle}{{\langle \pi^+_{++} \pi^-_{++} \rangle}^2} = Z e^{-\delta E_0 \left| \tau \right|} + \text{excited-state contributions},
\label{eq:CorrRatio}
\end{equation}
which may be expanded as
\begin{equation}
R(\tau) = Z \left[1 - \left| \tau \right| \delta E_0 + \frac{\tau^2}{2} (\delta E_0)^2 + …\right]+...
\label{eq:ratio}
\end{equation}
where $Z$ is some positive real constant. 
Here, for simplicity, we are assuming that the pions in the denominator of $R(\tau)$,
and two-pion system propagating in the numerator, are at rest,
as in the calculation of Ref.~\cite{Liu:2016cba}.\footnote{%
We note that Ref.~\cite{Liu:2016cba} does not fit to the ratio of correlators,
but rather to the numerator itself. This does not affect our argument as it is only
the part of the correlator dependent on $\delta E_0$ that is impacted by partial quenching.
}

For theories in which $s$-channel unitarity does not hold, 
the coefficient of the term quadratic in $\tau $
will not be half the square of the coefficient of $\left| \tau \right|$.
Explicit examples of this failure are given in Refs.~\cite{\HSWChPT,\DSPQLuscher}.
For the examples of interest here, our results in Sec.~\ref{sec:loop} show
that the expansion of the ratio $R(\tau)$ for $I=1$ and $I=0$ OS pions 
will not satisfy Eq.~(\ref{eq:ratio}).
If one nevertheless fits to this form, 
then an additional systematic error will appear in $\delta E_0$.

To estimate the size of this error, we need to know the relative contribution of
the linear and quadratic terms for the values of $\tau$ that contribute to the fit.
In the calculation of Ref.~\cite{Liu:2016cba}, and focusing on the lightest quark mass,
the energy shift is $\delta E_0  \sim -0.01 \; a^{-1}$,
while the fit extends up to $\tau \sim 10 \;a$.
Thus $\left| \tau \; \delta E_0 \right| \lesssim 0.1$ in the fit range,
implying that the quadratic term is a no larger than a $5\%$ correction to the linear term.
We therefore expect the error in the extraction of the energy shifts arising from the
breakdown of unitarity to be small, and in particular much smaller than the second
source of error we now discuss.

The second problem that arises from the breakdown of $s$-channel unitarity
concerns the quantization condition itself.
This converts energy shifts into scattering phase shifts. For our considerations it
will be sufficient to use the threshold expansion, which reads~\cite{\Luscher}
\begin{equation}
\delta E_0 = - \frac{4\pi a_0}{M_\pi L^3} 
\left[ 1 + c_1 \frac{a_0}{L} + c_2 \frac{a_0^2}{L^2} +  \cO(L^{-3}) \right]\,,
\label{eq:QCth}
\end{equation}
with $c_1 = -2.837$ and $c_2 = 6.375$.
If $s$-channel unitarity is violated, then only the leading order term is correct:
the $c_1/L$ correction, and all higher-order terms, are invalid.
To estimate the size of the error that arises from nevertheless
applying expression (\ref{eq:QCth}), we can determine the size of the
first two corrections for the parameters used in Ref.~\cite{Liu:2016cba}.
For their lightest quark mass, the OS pion mass is $aM_\pi=0.11985$,  on a lattice
of size $L/a=48$, and the scattering length obtained from solving the quantization
condition for $I=0$ OS pions is $Ma_0=0.73$.
The $1/L$ and $1/L^2$ corrections are then of size
$-36\%$ and $+10\%$, respectively,  compared to the leading term.
We therefore estimate the error 
%Note that the size of the corrections does not depend significantly on the ensemble chosen to analyze. 
arising from applying the quantization condition 
in a situation where it is not valid to be approximately $25\%$.

In principle, one could account for this source of error by calculating
the explicit form of the $1/L$ (and possibly higher order) corrections using \chpt\ at finite volume.
One would forgo the L\"uscher formalism and instead fit 
the finite-volume correlation functions that are calculated on the lattice
directly to the predictions from \chpt.
This approach was first proposed for the quenched theory~\cite{Bernard:1995ez},
and was worked out for Wilson fermions (without twisting) in Ref.~\cite{\HSWChPT}.
Unlike the two-particle quantization condition, it is not model-independent, 
but relies on the effective theory, here PQTM\chpt, to capture the
violations of unitarity.

In the present case, however, this approach is likely to be very difficult to implement.
This is because the one-loop contributions that lead both to the $\tau^2$ term
in Eq.~(\ref{eq:ratio}) and the $c_1/L$ term in Eq.~(\ref{eq:QCth}) involve, in part,
contributions with an OS $\pi^0$ and a sea-quark $\pi^0$ appearing in the loop,
and also contributions with two sea-quark $\pi^0$s,
as can be seen from Eq.~(\ref{eq:IZeroLoop}).
Both of these intermediate states are lighter than that with two OS pions,
so the simple exponential fall off assumed in Eq.~(\ref{eq:ratio}) is not valid,
and the correct form with which to replace Eq.~(\ref{eq:QCth}) is unclear.

Based on this discussion, we conclude that an additional systematic error of $\sim 25\%$ from
unitarity violation 
should be added to the result for $a_0$ obtained in Ref.~\cite{Liu:2016cba}.
We view this as an essentially irreducible error, due to the difficultly in correcting for
 violations of unitarity.
In addition, based on the analysis of Sec.~\ref{sec:LOamp}, we expect that there
is an $\cO(a^2)$ discretization error that could be as large as $100\%$. 
The key point here is that
we do not know whether the two $\cO(a^2)$ terms in Eq.~(\ref{eq:a0I0})---those
proportional to $\Delta_{\rm OS}$ and $w'_7$---partly cancel or not.
This is because we have, at present, no information on the size of $w'_7$.
However, as noted in Sec.~\ref{sec:LOamp}, calculations of the scattering
length for $I=2$ OS pions, which are not affected by unitarity violations, could be
used to determine $w'_7$ and thus correct for the LO discretization errors.
Thus this nominally very large discretization
error could be substantially reduced, very likely well below the size of the error due to
the violation of unitarity.

\section{Conclusion}
\label{sec:conc}

Twisted-mass fermions offer many advantages for computations in LQCD,
including automatic $\cO(a)$ improvement,
but these come at the price of the violation
of isospin and parity symmetries away from the continuum limit.
The size of unphysical effects due to the violation of these symmetries is,
generically, of $\cO(a^2)$, which is, for presently accessible values of $a$, 
of comparable size to physical contributions proportional
to the light quark masses. This makes it challenging to calculate quantities that are themselves
proportional to the light quark masses, an important example being the pion scattering amplitudes
near threshold.
While the $I=2$ amplitude evades this concern, as all discretization errors turn out to be
proportional to $a^2m_q$~\cite{Buchoff:2008hh},
the $\cO(a^2)$ isospin breaking makes it difficult to calculate the $I=0$ amplitude.

This problem was avoided in Ref.~\cite{Liu:2016cba} 
by working in a PQ setup using OS valence quarks, such that a modified version of isospin
was an exact symmetry. This allowed a result for the $s$-wave, $I=0$ scattering length
to be determined.
In this work we have shown, however,  that there are
two additional source of systematic error in this result that have not previously been accounted for.
The first is that OS pion scattering amplitudes do contain $\cO(a^2)$ contributions,
and these cannot be subtracted using previously-determined quantities.
Thus there is, {\em a priori}, a systematic error of $\cO(100\%)$ in the resulting scattering length.
We show, however, that this error can be substantially reduced if one also calculates the
scattering of $I=2$ OS pions, for a range of quark masses.
Such a calculation, though not previously undertaken, is much
simpler than that required for $I=0$ OS pions.

The second, and ultimately more challenging, source of error is due to partial quenching.
As we have recently emphasized, one cannot, in general, apply the two-particle quantization 
condition to PQ correlators~\cite{\DSPQLuscher}. What we have shown here is that this
prohibition applies for systems of two OS pions with $I=0$ (or $I=1$), whereas the quantization
condition can be used for the $I=2$ OS pion channel. The breakdown for $I=0$ implies
that, if one nevertheless proceeds using the quantization condition, then there is an additional
systematic error. We have estimated this error to
be $\sim 25\%$, and argued that it is essentially irreducible.

Our work suggests that a systematically improvable calculation of the $I=0$ scattering
amplitude near threshold using twisted-mass fermions is very challenging.
It appears to us that one must work in an unquenched setup,
so that the two-particle quantization condition is valid, 
and thus deal with the isospin breaking, which in effect makes this a problem involving
the $\pi^+_{\rm SS}\pi^-_{\rm SS}$ and $\pi^0_{\rm SS}\pi^0_{\rm SS}$ channels.
To remove the unphysical $\cO(a^2)$ effects from the results would likely require
using the \chpt\ results of Ref.~\cite{Buchoff:2008hh}, together with higher-order corrections.

\section*{Acknowledgments}

We thank Andr\'e Walker-Loud and Carsten Urbach for helpful comments and correspondence.
This work is supported in part by U.S. Department of Energy Contract No. DE-SC0011637.

\appendix

\section{Higher-order contributions in \chpt}
\label{sec:NLO}

In this appendix we collect some technical comments related to the \chpt\
calculations presented in the main text. 

First, we write out the NLO terms in the PQ chiral Lagrangian, which are used in Sec.~\ref{sec:Mixing}.
These consist of $a p^2$, $a m_q$ and $a^3$ terms,
\begin{equation}
\begin{aligned}
\mathcal{L}_{NLO} & = \hat a W_{4} {\rm str}\! \left[(\partial_{\mu} \Sigma) (\partial_{\mu} \Sigma^{\dagger})\right] {\rm str}\!\left[ \Sigma + \Sigma^{\dagger} \right]
+ \hat a W_{5} {\rm str}\!\left[ (\partial_{\mu} \Sigma) (\partial_{\mu} \Sigma^{\dagger}) (\Sigma + \Sigma^{\dagger}) \right] \\
& \quad - \hat a W_{6} {\rm str}\!\left[ \chi'^{\dagger} \Sigma + \Sigma^{\dagger} \chi' \right]
 {\rm str}\!\left[ \Sigma + \Sigma^{\dagger} \right]
- \hat a W_{7} {\rm str}\!\left[ 
\chi'^{\dagger} \Sigma - \Sigma^{\dagger} \chi' \right]
 {\rm str}\!\left[ \Sigma - \Sigma^{\dagger} \right] \\
& \quad - \hat a W_{8} {\rm str}\!\left[ \chi'^{\dagger} \Sigma^2 + \Sigma^{\dagger 2} \chi' \right] 
+ a^3 \cL_{\rm a^{3}}
\end{aligned}
\label{eq:LNLOW}
\end{equation}
where the notation is that of Ref.~\cite{\HSWChPT}, 
while the $a^3$ terms have the schematic form
\begin{equation}
\cL_{\rm a^{3}} \sim {\rm str}\! \left[\Sigma+\Sigma^{\dagger}\right]
+{\rm str}\! \left[\Sigma^3+\Sigma^{\dagger3}\right]
+{\rm str}\! \left[\Sigma^2\pm\Sigma^{\dagger2}\right]{\rm str}\! \left[\Sigma\pm\Sigma^{\dagger}\right]
+{\rm str}\! \left[\Sigma\pm\Sigma^{\dagger}\right]^2{\rm str}\! \left[\Sigma+\Sigma^{\dagger}\right]\,,
\label{eq:La3}
\end{equation}
where the $\pm$ signs in the third term are both positive or both negative.
Here, $\sim$ indicates that there is 
an independent LEC multiplying each term on the right-hand side.
We do not keep track of these LECs as they are not required for our analysis.

Second, we discuss why, when working at maximal twist, 
there are no terms involving odd powers of the field $\Pi$
in the expansion of the LO Lagrangian, Eq.~(\ref{eq:LchptLO}).
For the terms that are independent of $a$, this follows because, at maximal
twist, the condensate cancels the twist in the mass, leading to the standard kinetic
and mass terms expressed in terms of $\Sigma_{\rm ph}$, as shown by the first two terms in
Eq.~(\ref{eq:chlag}). These terms are symmetric under $\Pi \to -\Pi$, so that only
even powers of $\Pi$ can appear in their expansions.
For the terms induced by the nonzero lattice spacing, 
the key result is that they are proportional to $a^2$,
as the term linear in $a$ has been absorbed by a redefinition of $\chi$~\cite{SS}.
This implies that, when expressed in terms of $\Sigma_{\rm ph}$, as in Eq.~(\ref{eq:chlag}),
they involve two powers of $\Sigma_0\Sigma_{\rm ph}$.
These can appear in one or two straces, with each strace containing either
a hermitian or anti-hermitian combination, although the total term must be hermitian.
If one expands $\Sigma_{\rm ph}$, a given strace will thus have the form
\begin{equation}
i^{(x_1+y_1+x_2+y_2)} {\rm str}\!\left[ 
\Pi^{x_1} \: (\tau_3^{VS})^{y_1} \: \Pi^{x_2} \: (\tau_3^{VS})^{y_2} \right] \pm {\rm h.c.}\,,
\end{equation}
where the $x_i$ are non-negative integers, while the $y_i$ are either 0 or 1,
$h.c.$ indicates hermitian conjugate,
and we have used the result that $\Sigma_0=i \tau_3^{VS}$ at maximal twist.
Using the hermiticity of $\Pi$ and $\tau_3^{VS}$, together with the cyclicity of the strace,
we see that the hermitian conjugate term has the same (opposite) sign if
$N \equiv x_1+x_2+y_1+y_2$ is even (odd). 
Thus to avoid a cancellation $N$ must be even (odd) if the hermitian conjugate is added (subtracted).

Now we consider the ways in which one might obtain a cubic vertex from the $a^2$ terms.
If there is a single strace, it must contain an odd power of $\Pi$, so that $x_1+x_2$ is odd,
while we know that $y_1+y_2=2$ since there are two powers of $\Sigma_0$.
Thus $N$ must be odd. However, as there is a single strace, the term must be hermitian, and so,
from above, we know that terms with odd $N$ vanish. So there is no such vertex.

If there are two straces, to obtain a cubic vertex one
of them must contain an odd power of $\Pi$, and the other an even power.
Since $\sum_i y_i=1$ for both terms, we know that $N$ is odd for one strace, and even for the other.
Since the two terms are both either hermitian or anti-hermitian, this in turn implies
that one of the two terms vanishes.

Similar arguments can be used to demonstrate that the quadratic and quartic terms
arising from the NLO Lagrangian in Eq.~(\ref{eq:LNLOW}), vanish.

\bibliography{ref}

\end{document}